\documentclass[preprint,superscriptaddress, showpacs, floatfix, nobalancelastpage]{revtex4}
\usepackage{amsfonts}
\usepackage{amssymb}
\usepackage[dvips]{graphicx}
\usepackage{color}
\usepackage{amsmath}
\usepackage[bookmarksnumbered, bookmarks, breaklinks, linktocpage]{hyperref}

\begin{document}

\title{Magnetization, Magnetic Susceptibility and ESR in $\text{Tb}_{3}\text{Ga}_{5}\text{O}_{12}$}
\author{U.~L\"ow}
\affiliation{Theoretische Physik II, Technische Universit\"at Dortmund,
44227 Dortmund, Germany}
\author{S.A.~Zvyagin } 
\author{M.~Ozerov}
\affiliation{Dresden High Magnetic Field Laboratory (HLD),
Helmholtz-Zentrum Dresden-Rossendorf (HZDR), 01314 Dresden, Germany}
\author{U.~Schaufuss} 
\author{V.~Kataev} 
\affiliation{Leibniz Institute for Solid  State  and Materials Research IFW Dresden, D-01171 Dresden, Germany}
\author{ B.~Wolf}
\author{B.~L\"uthi}
\affiliation{
Physikalisches Institut, Universitat Frankfurt, Max-von-Laue Strasse 1, 60438
Frankfurt, Germany}

\begin{abstract}
We report on the measurement of the magnetic susceptibility
and of ESR transitions in the garnet substance  Tb$_3$Ga$_5$O$_{12}$ (TGG). 
The results are compared with a calculation
in the framework of crystal field theory  for the orthorhombic surroundings 
of the six inequivalent Tb ions of TGG. We also present a calculation of 
the magnetization for the three main crystal directions.
\end{abstract}

\maketitle

\section{Introduction }
\label{Intro} 
Tb$_3$Ga$_5$O$_{12}$ (TGG) is a dielectric material with a cubic garnet structure. 
Each Tb$^{3+}$ ion has the same $D_2$ symmetry in its own local coordinate system
which leads to pronounced crystal field (CEF) effects in various physical
properties. TGG exhibits a transition to a  antiferromagnetically ordered
phase at 0.35K (ref.\cite{Kama,Hamman}). 
The terbium ions are located on inter-penetrating 
triangular lattices, but they do not form a simple Kagome lattice, where the triangles lie in a 
plane, but are arrayed in three dimensions, forming a so-called Hyperkagome
lattice \cite{Ramirez}, which like the Kagome lattice is geometrically frustrated. 

Various interesting experiments have been performed recently with this
substance. 
One of them is the phonon Hall effect. 
In analogy to the classical Hall effect in conducting materials, the  appearance of a thermal
gradient in the direction perpendicular to both, the applied magnetic field
and the thermal flux, was observed \cite{Strohm2005,Inyushkin2007}. 
Another experiment is the acoustic Faraday effect where the rotation of a linearly
polarized ultrasonic wave in a magnetic field applied along the propagation
direction was observed and quantitatively analyzed \cite{us1,us2}. 
In another experiment the temperature dependence of the symmetry elastic
constants were measured 
and interpreted with a simple CEF-model \cite{araki}.

The aim of this paper is to show, that a CEF-Hamiltonian for the orthorhombic
sorroundings of the Tb ions is well fitted to describe the low energy spectrum
and the thermodynamic properties of TGG. We expect that this is also  
of intereset for a detailed model of transport phenomena like the
phonon Hall effect.

First we discuss the CEF-Hamiltonian,  next we calculate the
the magnetization in the three important crystal
directions in high magnetic field and compare the result with existing
experiments \cite{Guillot}. Afterwards we show experimental results of the magnetic
susceptibility and compare it with analogous calculations.
Finally we discuss our ESR results and interpret them using the same CEF-model.

\section{ CEF-Model }
\label{CEF} 

A Tb$^{3+}$ ion has 8 4f-electrons leading with Hund’s rule to S=3, L=3 and
J=6. The CEF for these ions with local orthorhombic symmetry can be described 
with  a Hamiltonian, introduced by Guillot 
et al. \cite{Guillot} in 1985, which in our notation reads:

\begin{eqnarray}
\label{H1}
{\cal H}_{loc}=&
                                        B_{20}   {\cal O}_{20}
                                       +B_{22}   {\cal O}_{22}
                                       +B_{40}   {\cal O}_{40}
                                       +B_{42}   {\cal O}_{42}
                                       +B_{44}   {\cal O}_{44}\\\nonumber
                                       +& B_{60}   {\cal O}_{60}
                                       +B_{62}   {\cal O}_{62}
                                       +B_{64}   {\cal O}_{64}
                                       +B_{66}   {\cal O}_{66}
+ g \mu_B \vec B \vec J.
\end{eqnarray}

Here $\vec B$ is the magnetic field in the local  coordinate 
system  of a Tb$^{3+}$ ion.
The ${\cal O}_{ij}$ are the Stevens operators as defined in \cite{Hutchings64},
and the $B_{ij}$ are the crystal field parameters.
The $B_{ij}$  can be calculated  from the 
$b_{ij}$ given in ref. \cite{Guillot}. We list the $b_{ij}$ from
\cite{Guillot} in table \ref{table1} and  give the explicit relation between
the $b_{ij}$ and the $B_{ij}$ in the appendix. For a detailed discussion of
the conventions used in defining the crystal field parameters see e.g.\cite{Wybourne}.

\begin{table}
\begin{center}
\begin{tabular}{|c|c|c|c|c|c|c|c|c|}
\hline
$b_{20}$ & 
$b_{22}$ & 
$b_{40}$ & 
$b_{42}$ & 
$b_{44}$ & 
$b_{60}$ & 
$b_{62}$ & 
$b_{64}$ & 
$b_{66}$ \\
\hline
  -81.0 & 
  169.0 & 
-2163.0 & 
  249.0 & 
  945.0 & 
  677.0 & 
 -155.0 & 
 1045.0 & 
   -4.0 \\
\hline
\end{tabular}
\end{center}
\caption{Crystal field parameters $b_{ij}$ for TGG in cm$^{-1}$ as given in ref.\cite{Guillot}}
\label{table1}
\end{table}

In this paper we use the Hamiltonian given by eq.\ref{H1} and take into account the six
inequivalent ions in the unit cell \cite{Levitin} to calculate the 
magnetization, the susceptibility and ESR matrix elements of TGG.
The local axis of the coordinate systems and the rotation matrices connecting 
the local and laboratory coordinate systems are given in table \ref{table2}.

\begin{table}
\begin{center}
\begin{tabular}{|c|c|c|c|}
\hline 
Rotation Matrix & $(e_x)_{local}$ & $(e_y)_{local}$ &$(e_z)_{local}$\\ 
   \hline 
$ R^1= 
\left(
\begin{array}{lll} 
    0                 &  \frac{1}{\sqrt 2 }                  & \frac{1}{\sqrt 2 }  \\
    0                 & - \frac{1}{\sqrt 2 }                 & \frac{1}{\sqrt 2 } \\
    1                 &   0                                  & 0
\end{array}
\right)$
                    & $[001]_c$ & $[1\bar{1}0]_c$  &$[110]_c$\\ 
   \hline 
$ R^{2}= 
\left(
\begin{array}{lll} 
    0 & - \frac{1}{\sqrt 2 }      &  \frac{1}{\sqrt 2 }  \\
    1 &   0                       &  0                \\
    0 &  \frac{1}{\sqrt 2 }       &   \frac{1}{\sqrt 2 } .
\end{array}
\right)$
                    & $[010]_c$ & $[\bar{1}01]_c$  &$[101]_c$ \\
   \hline 
                  
$ R^{3}= 
\left(
\begin{array}{lll} 
   1  &                     0  &   0 \\
   0  & \frac{1}{\sqrt 2}      &   \frac{1}{\sqrt 2}\\
   0  & -\frac{1}{\sqrt 2}     &   \frac{1}{\sqrt 2 } 
\end{array}
\right)$
       
             & $[100]_c$ & $[01\bar{1}]_c$  &$[011]_c$\\
   \hline 
$ R^4= 
\left(
\begin{array}{lll} 
    0 &   \frac{1}{\sqrt 2 } & -\frac{1}{\sqrt 2 } \\
    0 &  \frac{1}{\sqrt 2 } & \frac{1}{\sqrt 2 }\\
    1 &    0                & 0
\end{array}
\right)$
                    & $[001]_c$ & $[110]_c$  &$[\bar{1}10]_c$\\
   \hline 
$ R^{5}= 
\left(
\begin{array}{lll} 
    0  & \frac{1}{\sqrt 2 } & \frac{1}{\sqrt 2 } \\
    1  & 0                  &  0 \\
    0  & \frac{1}{\sqrt 2 } &  - \frac{1}{\sqrt 2 } .
\end{array}
\right)$
                            & $[010]_c$ & $[101]_c$  &$[10\bar{1}]_c$\\
   \hline 
                  
 $R^{6}=
 \left(
\begin{array}{lll} 
     1                & 0                   & 0 \\
   0& \frac{1}{\sqrt 2} &    -\frac{1}{\sqrt 2 } \\
   0& \frac{1}{\sqrt 2} &    \frac{1}{\sqrt 2 } 
\end{array}
\right)$

                     & $[100]_c$ & $[011]_c$  &  $[0 \bar{1}1]_c$\\ 
   \hline 

\end{tabular}
\end{center}
\caption{Rotation matrices from the local coordinate systems 
 of the six inequivalent Tb ions
to the laboratory system (column 1). Axis of  the local coordinate systems
 of the six ions (column 2,3,4).}
\label{table2}
\end{table}

\begin{figure} 
\begin{center}
\includegraphics[width=14cm,angle=0]{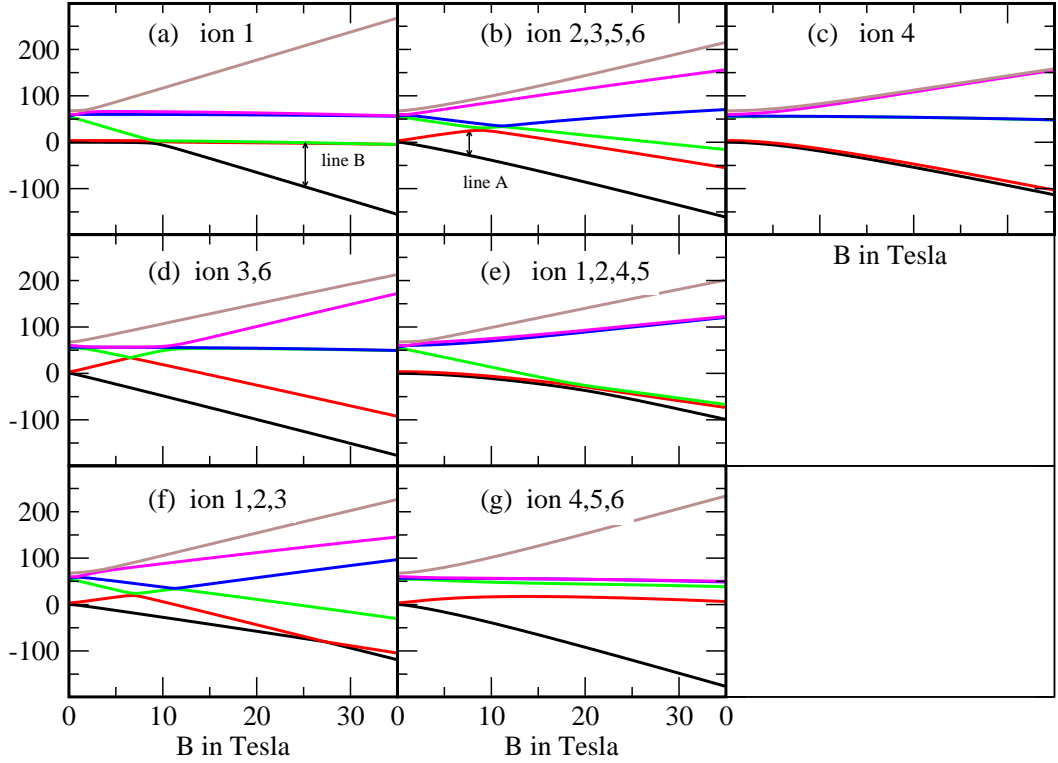}
\end{center}
\caption{Figure a,b,c show the six lowest energy levels for field in $[110]_c$
  direction, figure d,e are the corresponding levels for field in $[100]_c$ direction, and figure f,g 
  for field in $[111]_c$ direction. The corresponding ions are marked in the
  figures, the energies are in Kelvin and the
  magnetic field is in Tesla.} 
\label{fig2}
\end{figure}

Diagonalization of  the Hamiltonian eq.\ref{H1} gives the energy levels for the
 six ions as a function of magnetic field. These are given in fig.\ref{fig2}. For B=0
 we find a low lying quasi doublet, 3.7K apart, followed by a quasi triplet,
 with energy levels between 60K and 64K and a singlet at 67.8K. All other
 remaining singlets are above 433K. 
For cubic symmetry we would have a ground state doublet and an excited triplet 
followed by a singlet. The splitting from cubic to orthorhombic symmetry is therefore small.

In finite magnetic fields we observe
 crossovers of the energy levels. 
As field direction we take the three experimentally most
common cases, the magnetic field parallel to the cubic $[110]_c$ direction
(see ref. \cite{Guillot}), parallel to cubic $[100]_c$ direction (see ref.\cite{us1,us2})
and  parallel to cubic $[111]_c$ direction,
corresponding to the main axis of the crystal.
The most interesting crossovers occur for
 the lowest energy levels in fig.\ref{fig2}a at 9T for $B \parallel [110]_c$,
in fig.\ref{fig2}e at 20T for $B \parallel (100)$ and in
fig.\ref{fig2}f at 27T for $B \parallel [111]_c$.

\section{ Magnetization}

Since the energy levels of TGG display a complex structure of level crossings
 it is worthwhile to consider the magnetization, though due to
the strong magneto-caloric effect \cite{us2}, the magnetization of TGG  
in pulsed high fields is difficult to measure. 
We calculate 
\begin{eqnarray}
\langle J_k^{i} \rangle  
= -\frac{1}{Z} Tr \{ J_k \exp(- \beta ({\cal H}^{i}+g \mu_B \vec B \vec J))\}
\end{eqnarray}

where $\langle J_k^{i} \rangle$ is the contribution of the ith ion to the kth component $(k=x,y,z)$ of
the  magnetization in the laboratory system. The components of the total magnetization vector are given by $\langle J_k \rangle =
\sum_{i=1}^{6} \langle J_k^{i} \rangle $. Z is the partition function.
Note that the ${\cal H}^{i}$ are the Hamiltonians in the laboratory system
and also $\vec B$ is now the magnetic field in the laboratory system.
We obtained the ${\cal H}^{i}$ by a rotation of ${\cal H}_{loc}$ from 
the local systems $l_i$ \cite{Edmonds}\cite{Lind} . 
Equivalently one can of course calculate the magnetizations in the local systems
and transform them back to the laboratory system, which for a simple vector
quantity like the magnetization would be definitely the method of choice. However in view
of a more complex study in future concerning the elastic constants of TGG in
high magnetic field  we
choose to rotate the Hamiltonians. This means that the ${\cal H}^{i}$ we are dealing with
in our calculations are  more complex than the Hamiltonian given in
eq.\ref{H1}, since e.g., also ${\cal O}_{ij}$ with j odd appear. The
crystal field constants in the laboratory system were calculated by Mathematica.
Once these rotations are done the calculation of the oberservables discussed in
this paper can all be conveniently performed in the laboratory system.

\begin{figure} 
\begin{center}
\includegraphics[width=12cm,angle=0]{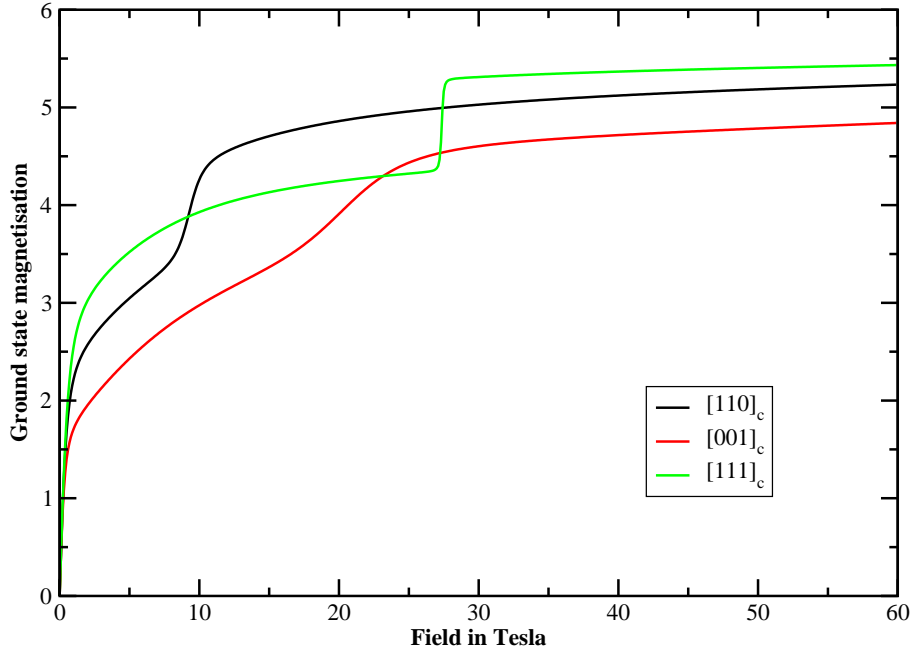}
\end{center}
\caption{Ground state magnetization $ |\langle \vec J \rangle| =(\langle J_x \rangle^2 + \langle
 J_y \rangle ^2 + \langle J_z \rangle^2)^{1/2}$ 
 with the field parallel to the $[110]_c, [001]_c, [111]_c$. The saturation is
 at $ |\langle \vec J \rangle|  =6$.}
\label{fig_mag}
\end{figure}

Fig.\ref{fig_mag} shows the total magnetization 
$ |\langle \vec J \rangle| =(\langle J_x \rangle^2 + \langle J_y \rangle ^2 + \langle J_z \rangle^2)^{1/2}$ at T=0 for the three
field directions. 
The steps and plateaux in the magnetization mark  cross-overs of the
lowest levels as pointed out above. 
Changes of slope also occur because of the summation of the contributions
of the different ions. So far only the magnetization in $[110]_c$ direction was
measured and calculated. 
For this direction our results agree with Guillot et al. \cite{Guillot}.
We want to point out, that the magnetizatons in the other directions also
exhibit interesting cross-over phenomena, which could be used to corroborate
the validity of the underlying CEF-model up to high fields.
Crossover effects at 20T for $B \parallel [100]_c$ have been observed
previously in acoustic measurements \cite{us2}.

Since in ref.\cite{Guillot} the magnetization was given in units 
of $\frac{\mu_\text{B}}{\text{molecule}}$ we make a quantitative comparison with our calculation. 
We get from ref.\cite{Guillot} for $\text{B}=15\text{T}$ an experimental value of 
$\approx  41.5 \frac{\mu_\text{B}}{\text{molecule}} $ for 2 TGG molecules
and from our  calculation $|\langle \vec J \rangle| _{[110]}  = 4.72/\text{(Tb ion)}$. 
This gives for 6 Tb ions and $g_L=1.5$ a value of $42.5 \frac{\mu_\text{B}}{2 \text{molecules}}$
 in close agreement with the experiment of ref.\cite{Guillot}.

\section{Magnetic Susceptibility}

\begin{figure} 
\begin{center}
\includegraphics[width=10cm,angle=0]{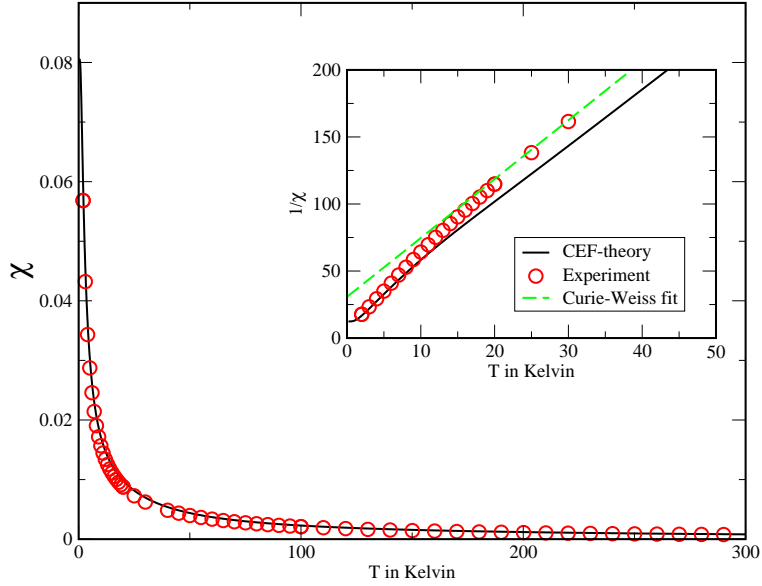}
\end{center}
\caption{Measured magnetic susceptibility  $\chi$ (red points) and
  result of CEF-calculation (black line) versus temperature in Kelvin. 
The inset shows the same data plotted as $1/\chi$
  compared with a Curie-Weiss fit (green dashed line).
 All the susceptibilites shown are dimensionless.}
\label{fig_sus}
\end{figure}

The magnetic susceptibility was measured in the temperature range between 
$2 \text{K} \leq \text{T} \leq 300 \text{K}$ and in a magnetic field of $ \text{B} = 0.1 \text{T} $ using a 
Quantum Design SQUID magnetometer. All measurements were performed on a single
crystal  with the external field oriented parallel to the cubic $[100]$
direction.  The data were corrected for the temperature-independent
diamagnetic core contribution  and the magnetic contribution of the sample
holder. The latter was determined from an independent measurement.
The experimental data versus temperature are shown in fig.\ref{fig_sus}.

A fit by a Curie-Weiss law
$\chi^{-1} = (\text{T}-\Theta)/\text{C}$ 
from  10K to 300K 
gives $\Theta = -7 \text{K}$ and  the Curie constant $\text{C} = 0.2277 \text{K}$. 
Evaluating  C  gives an effective magnetic moment of $\text{p}_\text{eff} =
9.25$ close to $\text{p}_\text{eff} = g_L \sqrt{J(J+1)} = 9.72$ for $\text{J}
= 6$ at high temperatures.
A $\theta = -7\text{K}$ compared with a N\'{e}el-temperature  $T_N =0.35\text{K}$ implies a strongly frustrated
spin system in the ordered anti-ferromagnetic region.

We also calculated the magnetic
susceptibility of TGG using the CEF scheme,
that is we  calculated  $\chi^i(T)$ of ion i in the laboratory system
from the free energy with the formula

 \begin{eqnarray}
\chi^i(T)=-\frac{\partial^2 F^i}{\partial B_z^2}  \nonumber  
=&\frac{1}{Z^2}\frac{1}{k_B T} \left( \sum_n \langle n| J_z | n\rangle
\exp{(\frac{-E_n}{k_BT})}\right)^2\\ \nonumber  
&-\frac{1}{Z} \frac{1}{k_B T}  \sum_n| \langle n| J_z | n\rangle|^2
\exp{(\frac{-E_n}{k_BT})}\\ 
&+\frac{2}{Z}   \sum_{n\neq m} \frac{|\langle n| J_z | m\rangle|^2}{E_n-E_m} \exp{(\frac{-E_n}{k_BT})}
\ \ \text{with} \ \  i=1,..6
\label{eq_sus}
\end{eqnarray}

and obtained the total susceptibility
$\chi(T)=\sum_{i=1}^6\chi^i(T)$ by summing 
over the contributions from the six inequivalent ions.
It is obvious that the magnetic susceptibility of
cubic TGG is isotropic. This can be also seen from fig.2 where the initial slopes
for the 3 different field directions are the same.
The calculated susceptibility is plotted as a function of temperature
together with the experimental data in fig.\ref{fig_sus} .
The measured $\chi_m$ in $\text{cm}^3/\text{Mol}$ TGG has been divided by the
molvolume using the parameters of table \ref{table3}. Without any fitting parameters the
agreement between experiment and calculation can be considered as very
good,except for a small deviation between 20K and 50K.

\begin{table}
\begin{center}
\begin{tabular}{|c|c|}
\hline
$\text{Density} $ & $\rho  = 7.22\ \text{g/cm}^3 $\\ 
  Molecular weight  & M = 1017.37      \\ 
Molvolume  & $V_M = 140.91 \text{cm}^3/\text{Mol}$ \\
 Number of Tb-ions per   $\text{cm}^3$ & $n =1.28 \times 10^{22}\text{cm}^{-3}$ \\
Land\'{e} g-factor  & $\text{g}_\text{L}=1.5$ \\
\hline
\end{tabular}
\end{center}
\caption{Physical parameters for TGG}
\label{table3}
\end{table}

As an example we compare the calculation with the experiment for $\text{T}=20\text{K}$.
From eq.\ref{eq_sus} we get that
$\frac{ d<J_i>}{dB} = 0.548 \text{T}^{-1}$ at $\text{T}=20\text{K}$.
With $\text{M}_i = \text{n g}_\text{L} \mu_\text{B} <J_i>$ and the values of table \ref{table3} this leads to $\chi_m
= 0.97 \times 10^{-2} \text{cm}^3/\text{cm}^3$ 
which has to be compared with the experimental value $\chi_\text{exp} = 0.88
\times 10^{-2} \text{cm}^3/\text{cm}^3$. 

\section{ESR Experiments}

The ESR  experiments were done employing a multiple-frequency 14 T  ESR
spectrometer \cite{Golze}
 equipped with a millimeter-wave  vector network analyzer (product
of AB Millimetre)  and a 16 T ESR  spectrometer equipped with VDI radiation
sources (product of Virginia Diodes Inc.)  
similar to that described in \cite{Zvyagin}.
The external field was applied parallel to the propagation direction of the
electromagnetic radiation  and with perpendicular incidence on the $[110]_c$ face 
of the TGG crystal. The electromagnetic field vectors $\vec e,\vec b$ are in free space
transverse waves and the Zeeman Hamiltonian reads 

\begin{eqnarray}
H_z  = g \mu_B \vec J \ \vec b_{rf}
\end{eqnarray}

\begin{figure} 
\begin{center}
\includegraphics[width=12cm,angle=0]{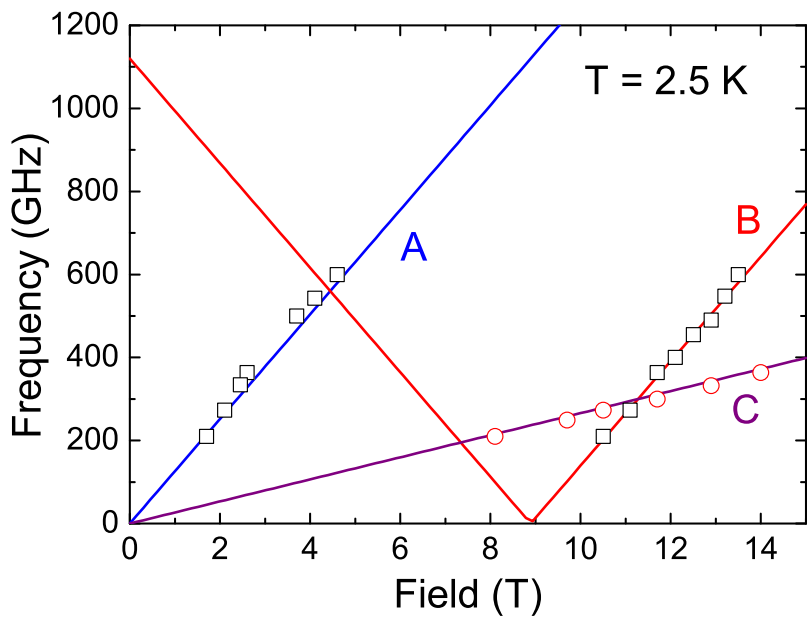}
\end{center}
\caption{ESR transitions measured at T=2.5K, with the field parallel to
the $[110]_c$ direction.The lines correspond to ESR  modes as described in the text.}
\label{fig1}
\end{figure}

\begin{figure}
\begin{center}
\includegraphics[width=14cm]{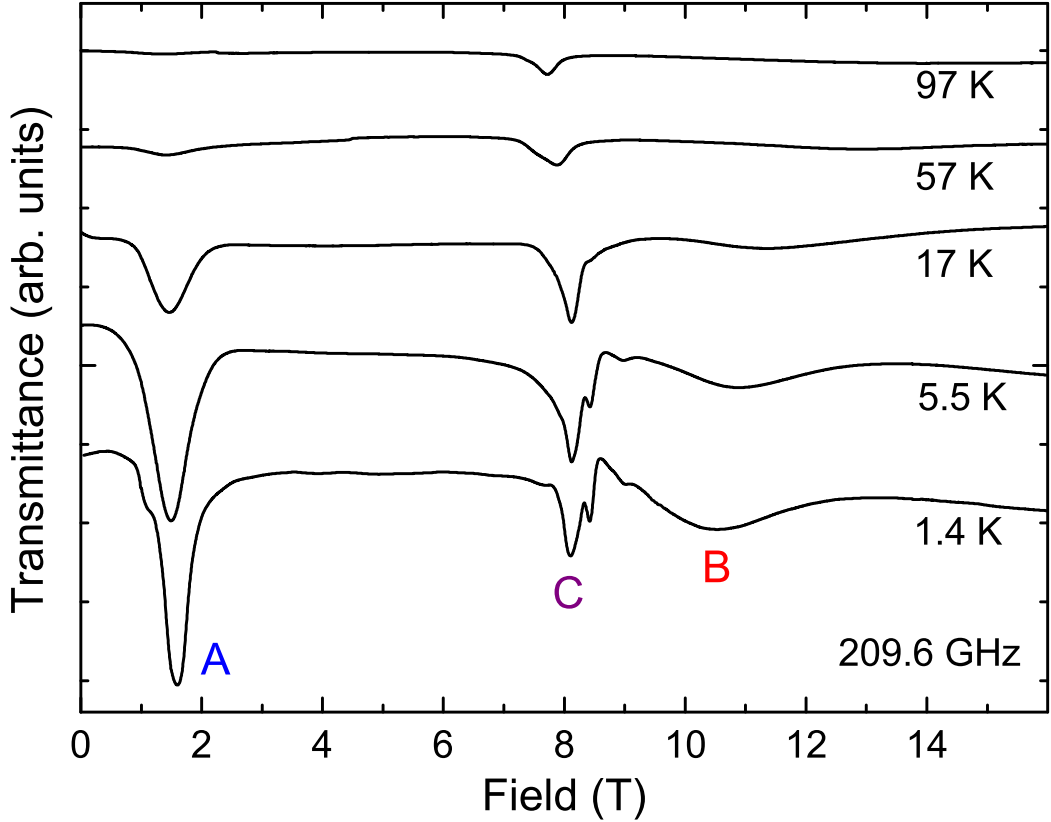}
\end{center}
\caption{Temperature dependence of the three observed ESR-lines A,B,C at a a frequency of 209.6 GHz. The field is
  applied in $[110]_c$ direction. Note, that the ESR spectra are offset for clarity.}
\label{fig_temp}
\end{figure}

with linear polarized waves and propagation
direction $\vec k \| [110]_c$, which in the local coordinate systems 
corresponds to the z-direction and the $(\sqrt{\frac{1}{2}}, \pm \frac{1}{2}
,\frac{1}{2}) $  directions respectively. We performed the  measurements in the frequency range 200-600 GHz. 

Experimentally we found 3 ESR lines, labelled A,B and C.
Fig.\ref{fig1} shows the resonances we found at 2.5K and fig.\ref{fig_temp} gives the temperature
dependence of the resonance lines up to 97 K.
Trying to identify these lines within our calculated level scheme  
(fig.1 a,b,c) we find that there are only few transitions 
which we can expect to be detectable and which we actually have observed in ESR.

\begin{figure} 
\begin{center}
\includegraphics[width=12cm,angle=0]{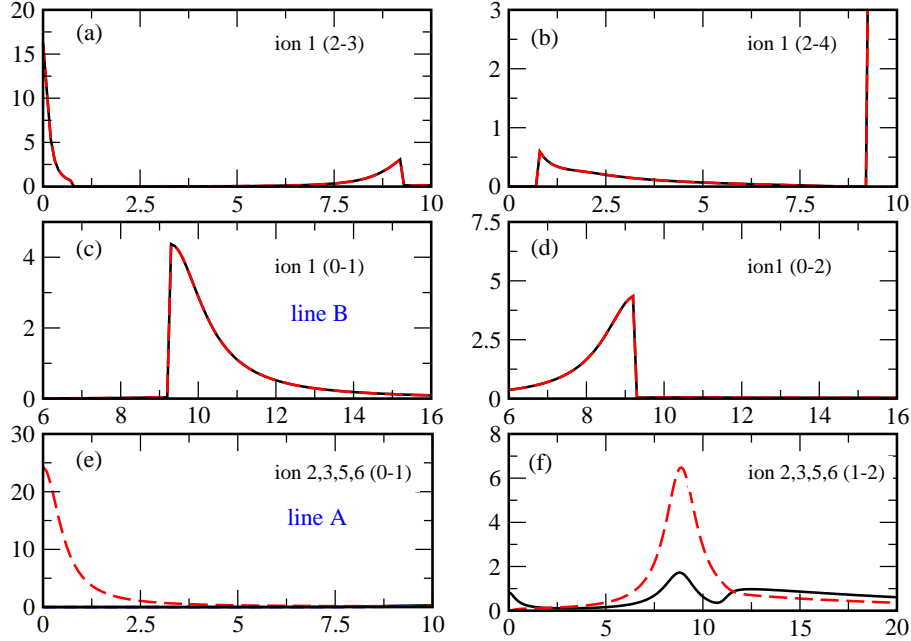}
\end{center}
\caption{Calcluated ESR transitions matrix elements 
$|\langle m|J_i|n \rangle|^2$, Where i=x is depicted as black line and i=y as red
  dashed line. The abscissa is always the magnetic field in Tesla.}
\label{fig_esr}
\end{figure}

To ensure that the  transitions we consider have nonvanishing weight
we also calculated the ESR transition matrix elements 
$\langle m|J_x|n\rangle$
and $\langle m|J_{y}|n \rangle$ between levels n,m. 
The relevant transition probabilities are shown in the fig.~\ref{fig_esr}. 
Since however the microwave intensities used in the experiment are strongly dependent
on the frequency we can only make a qualitative comparison with the calculated
matrix elements.

In detail we considered the following transitions in the level scheme 
as possible candidates for the experimentally observed lines A,B and C:
For the ions of type 1 (fig 1a) and magnetic field below 9 Tesla
the transition from the quasidegenerate doublet to the second state has
nonvanishing matrix elements (fig 6 d)
(note that we count the ground state as line 0, line 1 represents the first 
excited state, etc).
Also transitions from exited states i.e. between line 2 and line 3 or line 4
have nonvanishing weight (see figure 6 a and b).
For ions of type 1 and magnetic field larger than 9 Tesla we expect that 
the transition between the ground state and line 1 or 2 should be detectable
(fig. 6c), however the transition (0-2) has vanishing weight in this
field range (fig. 6d). 

For the ions of type 2,3,5,6 (fig 1b) 
there are two possible ESR-transtions:
For magnetic field below 9 Tesla the transition $0\leftrightarrow 1$  (see
fig. 6e) and the transition $1\leftrightarrow 2$ (see fig.6f).

For ions of type 4 (fig 1c) we do not expect to see any transitions in ESR.

Next comparing  the calculated level scheme with the experimental results  
shown in fig.~\ref{fig1} we come to the following conclusion:

Line A: From the frequency and temperature dependence we can conclude that we
deal here with a transition from the ground state. It is an increasing branch
and can be observed in the field region between $1.7$ to $4.7$ Tesla (see
fig.\ref{fig1}). 
It can be nicely fitted within the levels 0-1 for ions 2,3,5,6
(fig.\ref{fig2} and fig.\ref{fig_esr}) and gives, using $h \nu = g \mu_B B$
a g-factor 9.34 from the level scheme and from experiment we find a g-factor
of $g = 9$.  Above 9T one cannot follow this line further because the
corresponding energy levels (0-1) are constant and the matrix elements are
vanishing.

Line B: 
Also line B is a ground state excitation between the energy levels 0-1 
and 0-2 for ion 1 (fig.\ref{fig2} a) but observed in experiment only for $\text{B}>10.5\text{T}$.
Also the corresponding matrix element  $\langle 0 |J_{x,y}| 1 \rangle$  is
vanishing below 9T (fig.\ref{fig_esr}) which thus correponds to our experimental finding.
Note that  matrix element $\langle 0 |J_y| 2\rangle$ has some weight for
magnetic field  close to 9T on both sides. 
Since this is a decreasing branch
we could observe such a transition only below 7.4T with our lowest microwave frequency of
200 GHz. But for these fields the matrix element becomes rather small. Therefore line B can only be
observed for $\text{B} > 9T$. The experiment gives data up to 14T despite the fact that the corresponding
matrix elements become rather small. This can be due to the fact that the microwave intensity is
rather large for the higher frequencies as mentioned above.
The g factor for the observed B-line is again g = 9
and from our calculation we find g=8.7.

Line C: Finally the observed line C has spectral weight above and below the
level transition of 9T as seen in fig.\ref{fig1}.
It has a g factor of g = 1.9 and cannot be fitted easily into the level scheme 
of fig.\ref{fig2}. Since it is visible already at low temperatures we conclude
that it cannot be a transition within excited states, in particular it cannot
be identified with the transition 1-2
for ions of type 2,3,5,6 which has a g-factor g=1.6. 

But there are other possibilities for line C. One obvious idea
is a resonance line due Tb$^{4+}$ impurities. 
It is known  that 
the garnet structure can have Tb$^{4+}$ with $\text{L} = 0$, $\text{S} = \text{J} = \frac{7}{2}$ like
Gd$^{3+}$. 
This could give a line with g = 2 as observed. 
By tilting the sample in the magnetic field by $10^\circ$ and $20^\circ$ respectively the line 
splitted up considerably  for the frequency $\nu = 96.8$
GHz. 
This should not occur for a S-state ion with complete isotropy.
But a splitting occurs for line A too.
Another possibility is a Tb$^{3+}$ion on octahedral site substituting 
a Ga$^{3+}$ ion (see ref.\cite{Inyushkin2007}).

Considering the energy level diagram for the six different Tb$^{3+}$ ions and
the  given frequency range at our disposal the following points should be
noted: We cannot observe the splitting of the quasi doublet with $ \Delta \text{E} =
3.7\text{K} = 77 \text{GHz}$. We have not observed other transitions from excited states 
although we took measurements up to temperatures of 100K. 
The observed resonances are all on branches with increasing field.

To summarize, we find that the two ESR-lines A and B shown in fig.\ref{fig1} correspond  
very well with our calculations. 
The effective g-factor is $g=9.34$ for resonance A for $B<9T$ and ions 2,3,5,6.  
For resonance B and $B>9T$ we find an effective g-factor $g=8.7$ for ion 1.
as compared to the experimental value of $g=9$ for both line A and line B.

We want to emphasize here, that the description 
by the model  given in Eq. 1 is of course limited. Also
the crystal-field parameters, which are usually determined at
low magnetic fields  might be slightly
field dependent and it is not clear to what extend the small
but nonzero exchange interaction modifies the energy levels.

\section{Summary}

Taking all the presented experiments together we can say that the CEF scheme,
introduced by Guillot et al \cite{Guillot}, can explain magnetization, magnetic
susceptibility and ESR lines quite well. It is rather satisfying that all
these widely different magnetic effects presented in this paper originate in a
complicated garnet structure with 6 Tb ions with different local symmetries, 
which can be quantitatively described.  
 In the dicussions for the magnetization M, the magnetic susceptibility $\chi$ and
         the ESR we have neglected interaction effects between the ions. The existence of
         $T_N$ and $\Theta$  point of course to interaction effects. But they are too small to be
         observed in the magnetization M and in ESR.

\section{Appendix}
\label{app} 
The $B_{ij}$ used in eq.\ref{H1} can be calculated form the $b_{ij}$ given
in table \ref{table1}  by the following equations:
\begin{eqnarray}
&B_{2j}=  \frac{b_{2j} \alpha_J}{f_{2j}} & \ \ j=0,2\\\nonumber
&B_{4j}=  \frac{b_{4j} \beta_J}{f_{4j}} & \ \ j=0,2,4\\\nonumber
&B_{6j}=  \frac{b_{6j} \gamma_J}{f_{6j}} & \ \ j=0,2,4,6,
\label{A2}
\end{eqnarray}

where
\begin{eqnarray}
\label{A3}
&\alpha_J=\frac{-1}{99}\\
&\beta_J=\frac{2}{16335}\\
&\gamma_J=\frac{-1}{891891}
\end{eqnarray}.

and the $f_{ij}$ are listed in table \ref{table4}.

\begin{table}
\begin{center}
\begin{tabular}{|c|c|c|c|c|c|c|c|c|}
\hline
$ f_{20}$ & 
$ f_{22}$ & 
$ f_{40}$ & 
$ f_{42}$ & 
$ f_{44}$ & 
$ f_{60}$ & 
$ f_{62}$ & 
$ f_{64}$ & 
$ f_{66}$  \\
 \hline 
$2$                   &
$\frac{2}{\sqrt{6}}$  &
$8$               &
$\frac{8}{\sqrt{40}}$     &
$\frac{8}{\sqrt{70}}$     &
$16         $     &
$\frac{16}{\sqrt{105}}$   &
$\frac{16}{\sqrt{126}}$   &
$\frac{16}{\sqrt{231}}$   \\
\hline 
\end{tabular}
\end{center}
\caption{ Coefficients $f_{ij}$}
\label{table4}
\end{table}

\acknowledgments
This work was partly supported by the Deutsche Forschungsgemeinschaft and
EuroMagNET II (EU contract No. 228043).

\end{document}